\documentclass{ptephy_v1}

\preprintnumber{XXXX-XXXX} 

\usepackage{anyfontsize}

\usepackage{amsmath}
\usepackage{amssymb}
\usepackage{amsfonts}
\usepackage{textcomp}
\usepackage{gensymb}
\usepackage{graphicx}
\usepackage{needspace}

\newcommand\apj{Astrophys. J.}
\newcommand\prd{Phys. Rev. D}

\AtBeginDocument{
  \label{CorrectFirstPageLabel}
  
}

\begin{document}

\title{The Structure of Magnetically Dominated Energy Extracting Black Hole Magnetospheres: Dependencies on Field Line Angular Velocity}


\author{Kevin Thoelecke}\affil{Department of Physics, Montana State University, Bozeman, Montana 59717, USA \email{kevin.thoelecke@montana.edu}}

\author{Masaaki Takahashi}\affil{Department of Physics and Astronomy, Aichi University of Education, Kariya, Aichi 448-8542, Japan}

\author[1,3]{Sachiko Tsuruta}\affil{Kavli IPMU (WPI), University of Tokyo, Kashiwa, Chiba 277-8583, Japan}


\begin{abstract}
If a magnetically dominated magnetosphere is to extract a black hole’s rotational energy and transmit it to distant regions, then the inner light surface of that magnetosphere must lie within the ergoregion.  That inner light surface condition limits the angular velocity of magnetic field lines.  We take the distribution of magnetic field line angular velocity on the horizon to be a useful proxy for inner light surface location and study how different distributions affect the structure of energy-extracting magnetospheres.  Within magnetospheres that exhibit differential field line bending towards both the azimuthal axis and the equatorial plane, we find that the total Poynting flux energy directed outward along the azimuthal axis can vary by over a factor of 100 for a single value of black hole spin.
\end{abstract}

\subjectindex{xxxx, xxx}

\maketitle


\section{Introduction}

Rotating black holes can carry enormous amounts of extractable energy; in 1971 Christodoulou and Ruffini \citep{ChristodoulouRuffini1971} calculated that on average $6\%$ of an uncharged black hole's mass could be extracted, peaking at $29\%$ in the limit of extreme rotation.  Blandford and Znajek \citep{BZ77} later demonstrated that electromagnetic braking through an appropriately configured magnetosphere can be a practical method of extracting that rotational energy and transmitting it to distant observers.  Ever since energy-extracting black hole magnetospheres have been cited as potential drivers of some of the most energetic astrophysical objects, ranging from gamma-ray bursts to active galactic nuclei.  

Within the context of stationary, axisymmetric, and ideal magnetohydrodynamics \citep{TNTT90} showed that for a net energy outflow to occur along an isolated magnetic field line the Alfv\'{e}n point of the plasma inflow along that magnetic field line must occur within the ergoregion.  This dependency on some type of ergoregion behavior is a common feature of mechanisms that extract black hole rotational energy.  The mechanical method proposed by Penrose \citep{Penrose1969,Penrose19692002Reprint}, for example, relies upon the creation of particles with negative energies via collisions or other processes within the ergoregion.

While the importance of the Alfv\'{e}n point of an isolated magnetic field line is known, the practical significance of the Alfv\'{e}n surface of a collection of magnetic field lines forming a complete magnetosphere is far less certain.  This is largely due to the analytic intractability of the equations governing plasma flows in curved spacetimes.  In our previous work \citep{TTT2017} we began to explore potential implications of Alfv\'{e}n surface location for near horizon magnetospheres by considering force-free magnetospheres as the limit of magnetically dominated magnetospheres.  In force-free magnetospheres the ingoing Alfv\'{e}n surface coincides with the inner light surface, the location of which is determined by the magnetosphere's rotational profile.

We found that relatively slowly rotating magnetospheres with inner light surfaces near the outer limits of the ergoregion resulted in the bending of magnetic field lines towards the azimuthal axis and the formation of jet-like structures. Relatively rapidly rotating magnetospheres with inner light surfaces near the horizon resulted in the bending of magnetic field lines towards the equatorial plane, compatible with a direct connection between the horizon and a disk or similar nearby accreting matter structure.  Due to the complex nature of the equations involved those results were arrived at numerically, but such tendencies can also be seen in a more restricted form analytically (Appendix \ref{App:PerturbedSolution}).

In our previous numerical work we made multiple assumptions \citep{TTT2017}, one being a uniformly rotating magnetosphere.  That assumption was convenient in that it allowed us to know a priori exactly where the inner light surface of a given magnetosphere would lie, and it allowed us to focus on zeroth-order effects of magnetosphere rotation.  Despite those conveniences, however, uniformly rotating magnetospheres are likely to be fairly crude approximations of real black hole magnetospheres.
   
The goal of this work was to relax the assumption of uniform magnetosphere rotation and study rotational profiles that might more closely correspond to astrophysical black hole magnetospheres.  Our primary interests are in near-horizon behaviors where the effects of a rotating spacetime are strongest, so we took the event horizon as a natural place to specify the rotation of a magnetosphere.  Specifically we chose to study distributions of field line angular velocity $\Omega_\text{F}$ on the horizon corresponding to the first two terms of a series expansion of an arbitrary distribution:
\begin{equation} \label{Eqn:Equation1}
\left. \Omega_\text{F} \right|_{r_\text{H}} = \left(A + B \sin \theta\right) \omega_\text{H}.
\end{equation}     
Here $\omega_\text{H}$ is the angular velocity of the horizon (i.e. the angular velocity of a zero angular momentum observer on the horizon), while $A$ and $B$ are unitless constants.  We have selected this form (the power of a sine) because of its symmetry across the poloidal plane and because powers of sines can be very useful for both analytic and numerical explorations in other contexts.  Field line angular velocity $\Omega_\text{F}$ corresponds to a magnetosphere's rotation in that it may be thought of as a measure of the rotational boost velocity to the plasma rest frame (an explicit definition of $\Omega_\text{F}$ may be found in Equation \ref{Eqn:OmegaF} of the next section).  Our previous work studied uniformly rotating magnetospheres with $A = [0 \ldots 1)$ and $B = 0$ for a full range of black hole spins and corresponding horizon angular velocities $\omega_\text{H}$.

In this work we studied $A = [0 \ldots 1)$ and $B = (-1 \ldots 1)$, as those values most completely encompass arbitrarily rotating energy-extracting black hole magnetospheres.  Values of $A$ less than 0 or greater than or equal to $1$ were not considered as they would not generally correspond to energy-extracting magnetospheres.  We applied the condition $0 \leq A + B < 1$ so that the magnetospheres would extract rotational energy along almost every magnetic field line (field lines along the azimuthal axis or corresponding to $A + B \sin \theta = 0$ being the few exceptions).   We focused exclusively on a spacetime with black hole spin parameter $a = 0.8m$, selected as being large enough to be interesting without being overly extreme and potentially less widely representative. 

From our previous work we expected (and found) that the most interesting $A$ and $B$ values would fall into a fairly narrow range corresponding to low field line angular velocities near the azimuthal axis and high field line angular velocities near the equatorial plane.  The primary reason we find such ranges to be more interesting is because they split a magnetosphere into two distinct regions purely as a function of magnetosphere rotation: a higher latitude region with field lines bending upwards towards the azimuthal axis and a lower latitude region with field lines bending downwards towards the equatorial plane.  Other distinct behaviors are possible (such as field lines at higher latitudes bending downwards and lower latitudes bending upwards), but the simultaneous natural formation of both collimated jet-like structures and structures reminiscent of horizon-disk structures might be of more astrophysical interest.

Many of the $A$ and $B$ pairs we studied are deliberately na\"{i}ve and might not be very relevant to plausible astrophysical contexts.  Nonetheless we still felt that their calculation was important.  Not only do they form a more complete set when viewing Equation \ref{Eqn:Equation1} as a generic expansion of arbitrary magnetosphere rotation, they also place more astrophysically relevant distributions of horizon field line angular velocity in a more complete context.  So while many of the magnetospheres we calculate are likely to be mostly mathematical curiosities, their illumination of potentially more interesting black hole magnetospheres still gives them value.     

In Section \ref{Sec:Results} we report on our results for all $A$ and $B$ pairs, but with greater emphasis placed on the pairs that are more likely to correspond to astrophysical black hole magnetospheres.  Before that in Section \ref{Sec:Background} we provide brief reviews of the assumptions we have made and the numerical techniques we have used.  In Section \ref{Sec:Discussion} we discuss some potential implications of our results before concluding.


\section{Assumptions and Numerical Techniques} \label{Sec:Background}

The primary difference between this work and our previous work \citep{TTT2017} is the relaxation of the condition of uniform field line angular velocity.  Therefore we will primarily provide summaries of the assumptions and numerical techniques used and direct those interested in more detailed discussions to our previous work.  


\subsection{Assumptions}


\subsubsection{Core Assumptions} \label{Sec:CoreAss}

We assume a black hole whose surrounding spacetime is adequately described by the Kerr metric in Boyer-Lindquist coordinates, corresponding to the line element:
\begin{equation}
ds^2 = \left(1 - \frac{2mr}{\Sigma} \right) dt^2 + \frac{4mar \sin^2 \theta}{\Sigma} dt d\phi - \frac{\Sigma}{\Delta} dr^2 \nonumber - \Sigma d\theta^2 - \frac{A \sin^2 \theta}{\Sigma} d\phi^2,
\end{equation}
where
\begin{align}
\Sigma &= r^2 + a^2 \cos^2 \theta ,\nonumber \\
\Delta &= r^2 - 2mr + a^2, 	\nonumber \\
A &= \left(r^2 + a^2 \right)^2 - \Delta a^2 \sin^2 \theta .
\end{align}
Throughout this work we apply a black hole spin parameter $a = 0.8m$.  We also assume that the black hole is surrounded by a perfectly conducting plasma that is both stationary and axisymmetric, with the axis of symmetry corresponding to that of the black hole.  We additionally take the magnetically dominated force-free limit and assume that plasma inertial effects may be discarded, such that the magnetosphere can be completely described by three parameters: the toroidal vector potential $A_\phi$, the field line angular velocity $\Omega_\text{F}$, and the toroidal magnetic field $\sqrt{-g} F^{\theta r}$.  The toroidal vector potential $A_\phi$ and the field line angular velocity $\Omega_\text{F}$ are related to the field strength tensor $F^{\alpha \beta}$ by:
\begin{gather} 
F_{\alpha \beta} = A_{\beta, \alpha} - A_{\alpha, \beta}, \nonumber \\
F_{r \phi} \Omega_\text{F} = F_{tr}, \nonumber \\
F_{\theta \phi} \Omega_\text{F} = F_{t \theta}. \label{Eqn:OmegaF}
\end{gather}    
Here a comma denotes a partial derivative.  The toroidal vector potential $A_\phi$ is conserved along magnetic field lines and as such is a useful flux function for the poloidal magnetic field.  The field line angular velocity and toroidal magnetic field are also conserved along magnetic field lines, respectively statements that field lines rotate rigidly and that energy and angular momentum Poynting fluxes are conserved.  The conservation of field line angular velocity is a consequence of stationarity, axisymmetry, and a perfectly conducting plasma (expressed as the vanishing contraction of the field strength tensor and its dual $\mathcal{F}^{\alpha \beta} F_{\alpha \beta} = 0$) and as such is also a conserved quantity when plasma inertial effects are considered.  The conservation of the toroidal field is the force-free limit of angular momentum flux conservation.  Further discussion of force-free conserved quantities may be found in \citep{BZ77}; further discussion of conserved quantities when plasma inertial effects are present may be found in \citep{BekensteinOron1978, Camenzind1986a, TNTT90}.        


\subsubsection{Boundary Conditions} \label{Sec:MonopoleCondition}


We assume reflection symmetry across the equator so that we need only solve for the structure of the magnetosphere in the upper half of the poloidal plane.  We then apply boundary conditions along the azimuthal axis and equatorial plane that are compatible with a ``monopolar'' magnetic field, in the sense that we assume fixed magnetic field lines tracing the azimuthal axis and equatorial plane (mathematically $A_\phi(\theta = 0) = A_{\phi \text{max}}$ and $A_\phi(\theta = \pi/2) = A_{\phi \text{min}}$ where $A_{\phi \text{max}}$ and $A_{\phi \text{min}}$ are fixed constants).  We use the term ``monopolar'' to describe those boundary conditions while noting that has the potential to be misleading.  There will often be a substantial toroidal component of the magnetic field, resulting in a helical magnetic field that only resembles a monopole when projected onto the poloidal plane.  Additionally, non-zero magnetic flux through a closed surface surrounding the black hole demands a reversal of the direction of the magnetic field below the equatorial plane, so ``split-monopolar'' would be a more appropriate term when the entire poloidal plane is considered.  In short ``monopolar'' should only be taken to describe the boundary conditions on the upper half of the poloidal plane where we conduct our numerical calculations, not the magnetosphere as a whole.

A field line tracing the azimuthal axis is not a significant restriction as stationarity and axisymmetry already imply a single magnetic field line extending straight upward from the pole.  However a single magnetic field line tracing the equatorial plane is a more severe restriction.  Although it might be physically reasonable in regions close to the horizon, the further away one gets the less reasonable it is likely to become.  This is because significant amounts of matter would generally be expected near the equatorial plane at least as close as the innermost stable circular orbit.  That matter would likely anchor many different magnetic field lines and be better described by higher order multipoles in the equatorial plane.

Despite their potential deficiencies, monopolar boundary conditions are still very useful in explorations of basic interactions between the electromagnetic fields and the background spacetime.  Other boundary conditions intrinsically assume a specific matter distribution outside the black hole and by extension a specific astrophysical context.  Any such assumed context not only introduces its own assumptions but also has the potential to introduce arbitrarily large forcings on the magnetosphere that might obscure interactions between the electromagnetic fields and the background spacetime.  In short the assumption of monopolar boundary conditions is a deliberate compromise; we are favoring a more fundamental exploration of black hole magnetospheres over direct applicability to any specific astrophysical context.


\subsubsection{Limited Domain} \label{Sec:LimitedDomain}


We do not extend our magnetospheres past the outer light surface (pulsar light cylinder analog) or $r = 20m$, whichever occurs sooner.  That choice is not made completely freely, as diffusive numerical techniques are generally incapable of finding magnetospheres that pass smoothly through both inner and outer light surfaces once a specific distribution of horizon field line angular velocity has been specified.  This is because the character of the equations involved changes across a light surface; the prefactor on the second derivatives of the vector potential changes sign.  This means that numerical schemes can fail to find many valid solutions, as for stability they must necessarily evolve the magnetosphere differently on either side of a light surface.  Finding solutions numerically then generally reduces to ``matching'' minimum energy solutions across a light surface by adjusting field-aligned conserved quantities.  Matching three regions across two light surfaces would require two different conserved quantities, but once a distribution of horizon field line angular velocity has been specified only the toroidal field remains as a free variable.  This means that numerical techniques will generally be incapable of finding matched solutions across both inner and outer light surfaces once a distribution of horizon field line angular velocity has been specified, even if such solutions exist.  

For example, within the monopolar boundary conditions we have set infinitely many solutions are known to exist that pass smoothly through both inner and outer light surfaces \citep{MenonDermer2007}.  However when numerical techniques are applied by matching across light surfaces only a single highly monopolar (minimum energy) solution is found \citep{CKP2013}.  As we are primarily interested in near-horizon behaviors in this work, we have chosen to limit ourselves to regions interior to the outer light surface.

Despite the fact that the outer light surface is a numerical limitation, our selection of a limited domain interior to $r = 20m$ or the outer light surface has physical motivations.  For example, we have assumed that the field-aligned conserved quantities are rigidly conserved across the entire magnetosphere, from the horizon to the outer boundary.  While the assumptions of stationarity, axisymmetry, and a force-free plasma that led to those conserved quantities might be approximately valid over any given region, as a field line grows in length small deviations from those assumptions can grow in significance.  Our limited domain can be thought of as the assumption that field-aligned quantities are only approximately conserved, and that near horizon values might differ significantly from more distant values.

Additionally, when the force-free limit is viewed as the magnetically-dominated limit of an ideal plasma flow, then consideration must be made of the ``separation surface'' between the inner and outer light surfaces that separates a plasma inflow from a plasma outflow (by demarcating the change in dominance from inward gravitational forces to outward centripetal forces).  Near the separation surface plasma effects can become significant and deviations from our assumptions can be expected, to some extent decoupling inner and outer magnetospheres near the separation surface.  We note that the separation surface is a suggestion, however, and not a rule - problematic effects (mathematically sourced by a diminishing Alfv\'{e}n Mach number and concurrent increase in plasma density) do not have to emerge there, but if they do not emerge at or interior to the separation surface they are guaranteed to emerge as the outer light surface is approached.  As such we are not concerned with addressing the numerical difficulties that emerge at the outer light surface, as in the problem space of ideal plasma flows we necessarily demand physical changes to the problem (from a plasma inflow to a plasma outflow) before arriving there.  Although in general we expect the separation surface to be a much more physically relevant indication of changing physics than the outer light surface, it is numerically trivial to extend the magnetosphere past the separation surface to the outer light surface, and we see no harm in doing so.

It is possible that the selection of the outer boundary might influence (or drive) the structure of the magnetosphere contained within it (e.g. spherical outer boundaries at $r = 3m$, $r = 4m$, and $r = 5m$ might result in different solutions).  We studied that behavior in initially developing the numerical techniques applied in our previous work \citep{TTT2017}, and found that such differences could indeed emerge if the treatment of the outer boundary was poorly implemented.  If done appropriately, however, the solutions obtained are identical and the outer boundary does not influence the solution.  In this work we continued to verify the apparent invariance of the solutions with outer boundary, and (in addition to other tests) calculated every magnetosphere a minimum of two times: once with a spherical outer boundary (typically between $r=3m$ and $r=4m$) completely interior to the outer light surface, and once with an outer boundary limited by the outer light surface (or $r = 20m$).  The solutions obtained were always identical, regardless of outer boundary.  For this reason we do not view the outer light surface as a ``boundary condition'', but rather the point at which the rigid application of stationary and axisymmetric force-free magnetohydrodynamics (extended from the horizon) has definitively broken down.  In a specific model that breakdown might be avoided or diminished by appealing to additional physics, which is a reason why \citep{BZ77} and others have discussed ``spark gaps'' and other plasma injection mechanisms.  A more generic exploration of the solution space of ingoing magnetospheres has no such luxury, however, and is restricted to a domain interior to the outer light surface \citep{MacDonaldThorne1982}.


\subsection{Numerical Techniques} \label{Sec:NumTech}


The only significant change to our numerical techniques from our previous work \citep{TTT2017} is our method of kink reduction across the inner light surface, which we have improved to allow for significantly reduced error levels there.  That error reduction does not modify our final results in any appreciable way but does allow for more efficient computation of magnetospheres.


\subsubsection{Magnetofrictional Method} \label{Sec:MFMethod}


To calculate magnetospheres we apply a relativistic extension of the magnetofrictional method developed by \citep{YSA1986}, similar to the method used by \citep{Uzdensky2005}.  This method takes an initial guess for the structure of a magnetosphere and then calculates the divergence of the stress energy tensor.  If that divergence does not vanish the configuration is invalid (or at least inconsistent).  To find  a valid configuration the invalid excess momentum fluxes are converted to the velocity $\mathbf{v}$ of a fictitious plasma via empirically determined ``friction''.  The magnetic fields $\mathbf{B}$ are then modified by the relativistic analog of the ideal induction equation $\partial_t \mathbf{B} = \mathbf{\nabla} \times (\mathbf{v} \times \mathbf{B})$.  The end result is that the vector potential $A_\phi$ is evolved via a simple advection equation until a solution is found:
\begin{equation}
A_{\phi, t} = - v^A A_{\phi, A}.
\end{equation}  
Here the uppercase Latin indices denote poloidal directions ($r$ and $\theta$) and the magnetofrictional velocity $v^A$ includes empirically determined weighting factors for convenience and stability as outlined in our previous work \citep{TTT2017}.  A demonstration that application of the magnetofrictional method will always result in a valid force-free magnetosphere under fairly general assumptions may also be found there. 

We have found that our numerical procedures always converge to the same solution regardless of the initial guess for the structure of the magnetosphere (initial $A_\phi$, $\Omega_\text{F}$, and $\sqrt{-g} F^{\theta r}$); poor initial guesses simply take longer to converge.  The magnetofrictional method is an energy-minimizing algorithm, so that single solution is at best an indication of the uniqueness of a minimum-energy state.  We believe it likely that in addition to that minimum energy solution there are also infinitely many magnetospheres compatible with our boundary conditions and assumptions that contain more energy and are therefore unstable in some fashion.  We therefore interpret the solutions found as being ``most compatible'' with the various assumptions made, but not unique.


\subsubsection{Kink Reduction}


Regions inside and outside the inner light surface are evolved using a different overall prefactor of $\pm 1$ on the magnetofrictional velocity in order to maintain numerical stability.  This causes a ``kink'' to develop across the inner light surface as the vector potential $A_\phi$ is evolved in opposite directions on either side.  To find a smooth solution we adjust the toroidal field as a function of the vector potential, $\sqrt{-g} F^{\theta r} (A_\phi)$, until that kink disappears.  To make modifications to that function we first shoot across the inner light surface from the inside (near horizon region) to the outside such that the near horizon region provides an inner boundary condition for the region outside the inner light surface.  We then modify the toroidal field corresponding to the shot grid squares by measuring how close to being force-free those squares are.  Specifically we use the error level of the shot grid squares (calculation of that error is discussed in the next section) to gradually correct the functional form of the toroidal field:
\begin{equation}
\left. \sqrt{-g} F^{\theta r} \right|_{\text{New}} = \left. \sqrt{-g} F^{\theta r} \right|_{\text{Old}} - \lambda \cdot \text{Error}.
\end{equation}
Here $\lambda$ is an empirically determined constant; optimal values vary widely, depending primarily upon grid resolution and current error level.

The above method of kink reduction differs slightly from our previous work \citep{TTT2017}.  Previously we adjusted the toroidal field by evolving both sides of the inner light surface separately, then measured the magnitude of the difference in $A_\phi$ across the light surface as an input in adjusting the toroidal field.  While that method works reasonably well, as the kink in $A_\phi$ becomes smaller it can become very difficult to accurately quantify and therefore reduce the error level of the final solution.

By directly using the error level to modify the toroidal field we are able to reduce the error along the inner light surface significantly from what was obtainable in our previous work. Error levels of at least $0.001\%$ are now fairly easy to obtain along the entire extent of the inner light surface (computation time being the primary limiting factor) while our previous method would sometimes struggle to significantly exceed $0.1\%$ as a worst case.  

Despite the advantages in error reduction, the primary motivation for the change in kink reduction method was to enable more rapid convergence to a solution.  Enhanced error reduction was merely a side effect that did not change our results in any appreciable fashion.

Both our previous and current methods for kink reduction are identical in basic principle to the method developed by \citep{Contopoulos1999} for pulsar magnetospheres and applied more recently by others \citep{CKP2013, NC2014,PanYuHuang2017} to black hole magnetospheres.


\subsubsection{Error Determination}


Our magnetospheres can contain both strongly monopolar regions as well as regions with very small current, so most commonly used measures of force-freeness that rely directly on some physical attribute of the fields (such as the ratio of Lorentz force and electric current, as in \citep{MG2004}) can yield unreliable results due to one or more of the measured physical attributes becoming vanishingly small.  We have therefore developed a more mathematical technique for measuring the error of our solutions.  A valid and self-consistent solution will have a stress energy tensor with vanishing divergence, so we measure how close to zero the divergence of a magnetosphere's stress energy tensor is in order to determine the magnetosphere's error level.  Specifically, we first separate the divergence into seven terms:
\begin{equation}
T^A{}^\beta{}_{; \beta} \sim \sum_{i = 1}^7 D_i = \delta.
\end{equation}  
The exact form of $D_i$ we use is detailed in our previous work \citep{TTT2017}; their sum is not completely equivalent to the divergence of the stress energy tensor because we apply overall weighting factors for convenience.  When $\delta$ is close enough to zero a solution has been found, close enough being determined by comparing $\delta$ to the largest of the $D_i$ terms:      
\begin{equation}
\left| \delta \right| < \epsilon \cdot \text{Max}\left(\left| D_i \right| \right).
\end{equation}  
We have set $\epsilon = 1\%$ over the entire domain as an adequate error level, but in practice most of the domain will be significantly less.  Typically the largest $r$ values are the last regions to achieve the $1\%$ level, at which time averages of $0.0001\%$ inside the ergosphere are common.  Magnetospheres with an error level of $10\%$ are generally not substantively different from those at $1\%$, which are in turn effectively indistinguishable from those at $0.1\%$ and below.  We chose $1\%$ as an error level in order to remove as much numerical uncertainty as possible while avoiding excessive amounts of computation time.


\subsubsection{Computational Specifics} 


The vector potential $A_\phi$ is calculated over a rectangular $(r, \theta)$ grid with 200 evenly spaced grid squares in $\theta$ and on average around 1000 variably spaced grid squares in $r$.  The radial spacing varies from magnetosphere to magnetosphere; magnetospheres with inner light surfaces near the horizon have tighter spacing there in order to adequately resolve the inner light surface.  The radial grid extends from just inside the horizon to $r = 20 m$ or the outer light surface, whichever is smaller, with radial spacing of around $0.1m$ near $r = 20m$.           

The toroidal field is implemented as a function of $A_\phi$ with over 1000 points of varied spacing between $A_{\phi \text{min}}$ on the equatorial plane and $A_{\phi \text{max}}$ on the azimuthal axis.  That spacing is determined by convenience, as convergence can be optimized by using very fine sampling in regions where the toroidal field as a function of $A_\phi$ is steep. 

To evolve the advection equation for $A_\phi$ we use an upwind differencing algorithm similar to the one described in \citep{HSW1984}.  One-sided finite difference approximations appropriate to that algorithm are made to evolve $A_\phi$, but centered finite difference approximations appropriate to the local grid spacing are used to determine the magnetofrictional velocity $v^A$.

The azimuthal axis and equatorial plane are taken as fixed boundaries.  At all other boundaries ($r_\text{min}$, $r_\text{max}$, and/or along the outer light surface) we shoot outwards using a quadratic fit after every time step, an approach that is largely equivalent to using one-sided derivatives on those boundaries.
 

\subsubsection{Performance}


We computed all magnetospheres on a single desktop computer with a 6-core Intel Haswell CPU assisted by Nvidia Kepler GPUs, which can generally find a magnetosphere at the $1\%$ error level within a few hours.  Exact time to completion can vary widely, though, from well under an hour to days in extreme cases.  Computation time is highly dependent upon how good the initial guess was, how tight the grid spacing is, and how optimal various empirical tunings (strength of friction, modification to the toroidal field, etc.) are.  At a high level our algorithm is conceptually similar to finding the root of a computationally expensive function by crawling along that function, and as such can be susceptible to large inefficiencies similar to those found when over or under evaluating an expensive function and from taking steps that are either too big or too small.  Significantly improving the speed of our code and its algorithmic inefficiencies should be possible, but we found the current performance level to be adequate for this work.


\section{Results} \label{Sec:Results}


We divide our results into three sections.  First we explore the general structure of the magnetospheres obtained as a function of field line angular velocity, measured by the $A$ and $B$ parameters in $\Omega_\text{F}(r_\text{H}, \theta) = (A + B \sin \theta) \omega_\text{H}$.  We then explore the rates of energy and angular momentum extraction from the black hole.  Lastly we explore the behavior of magnetospheres containing both jet-like regions and structures resembling horizon-disk connections in more detail, as any such obviously distinct regions allow for a finer exploration of magnetosphere properties. 

\begin{figure}[!h]
\centering
     \includegraphics[width=3.5in]{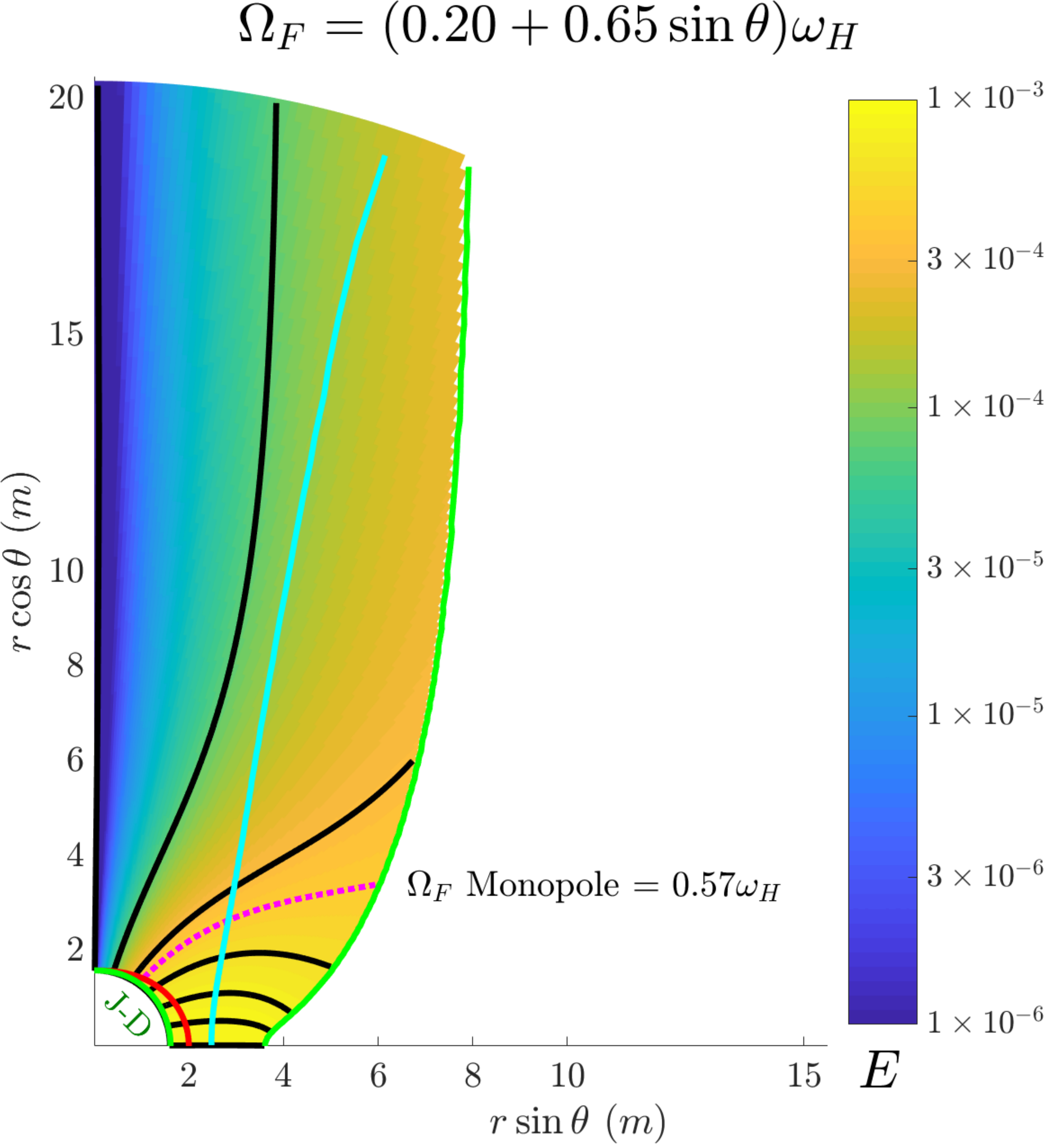}
     \caption{The structure of a magnetosphere with horizon field line angular velocity $\Omega_\text{F} = (0.20 + 0.65 \sin \theta) \omega_\text{H}$.  The black poloidal magnetic field lines are spaced evenly on the horizon.  The green lines trace the inner and outer light surfaces, the red line traces the boundary of the ergosphere, and the cyan line traces the separation surface (the point at which gravitational and centripetal forces are balanced).  The dotted magenta line traces the monopolar separatrix between field lines bending towards the axis and field lines bending towards the equatorial plane.  This magnetosphere is classified as a Jet-Disk magnetosphere, denoted by the ``J-D'' text inside the horizon.  The background shading denotes the magnitude of the conserved field-aligned Poynting flux; $E = (1/4\pi) \sqrt{-g} F^{\theta r} \Omega_\text{F}$.  A plot of this magnetosphere's ergoregion is shown in Figure \ref{Fig:Figure2}, and an additional 12 magnetospheres are displayed in similar fashion in Figure \ref{Fig:Figure3}.} 
	   \label{Fig:Figure1}
\end{figure}


\subsection{General Structure} \label{Sec:GeneralStructure}


We calculated 400 distinct magnetospheres with different horizon field line angular velocities $\Omega_\text{F} = (A + B \sin\theta) \omega_\text{H}$ using a spacing of $0.05$ in both $A$ and $B$ over the ranges $A = [0 \ldots 1)$ and $B = (-1 \ldots 1)$ under the condition that $0 \leq A + B < 1$.  It would be impractical to show all 400 in detail, so instead we classify magnetospheres based upon their general structure and show some representative types.  

\begin{figure}[!h]
\centering
     \includegraphics[width=3.5in]{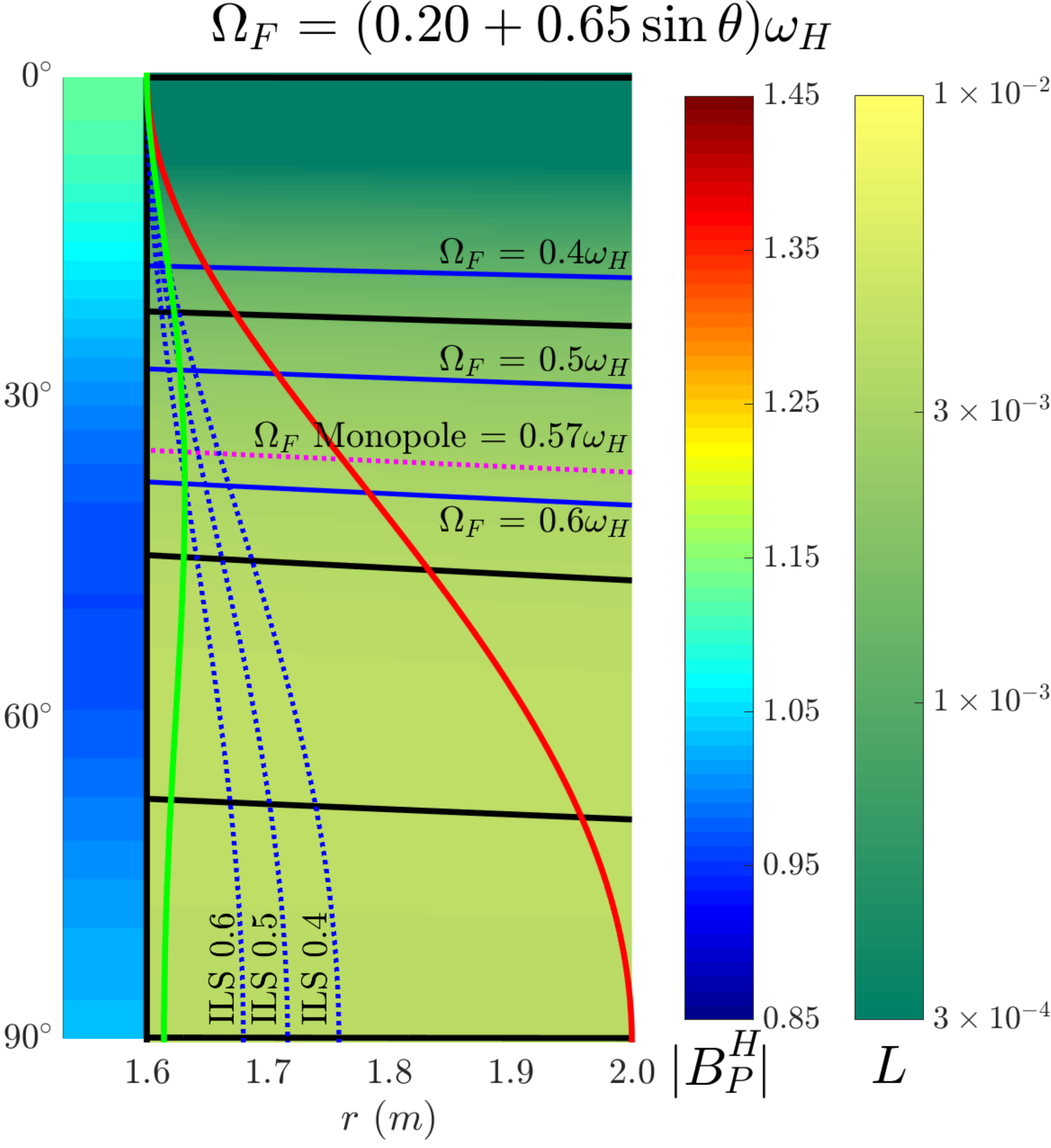}
     \caption{The structure of a magnetosphere with horizon field line angular velocity $\Omega_\text{F} = (0.20 + 0.65 \sin \theta) \omega_\text{H}$, the same as shown in Figure \ref{Fig:Figure1}.  The inner light surface is shown in green, the boundary of the ergosphere is marked in red, and the monopolar separatrix between Jet and Disk behaviors is shown as a dotted magenta line.  The black magnetic field lines are spaced evenly on the horizon.  The three blue magnetic field lines rotate with field line angular velocities $\Omega_\text{F} = 0.4 \omega_\text{H}$, $\Omega_\text{F} = 0.5 \omega_\text{H}$, and $\Omega_\text{F} = 0.6 \omega_\text{H}$.  The shading to the left of the horizon ($r_\text{H} = 1.6m$) is a measure of the poloidal magnetic field strength on the horizon; $|B_\text{p}^\text{H}| \sim A_{\phi, \theta} \csc \theta$.  The shading outside the horizon is a measure of both the toroidal magnetic field and conserved angular momentum Poynting flux; $L = (1/4\pi) \sqrt{-g} F^{\theta r}$.  The three dotted blue lines correspond to the inner light surfaces of uniformly rotating magnetospheres with $\Omega_\text{F} = 0.6 \omega_\text{H}$ (closest to the horizon), $\Omega_\text{F} = 0.5 \omega_\text{H}$, and $\Omega_\text{F} = 0.4 \omega_\text{H}$ (furthest from the horizon).  An additional 12 magnetospheres are displayed in similar fashion in Figure \ref{Fig:Figure3}.} 
	   \label{Fig:Figure2}
\end{figure}

In the poloidal plane there are only three things that a magnetic field line can do: bend upwards toward the azimuthal axis, remain straight, or bend downwards towards the equatorial plane.  We classify each of those three tendencies as being ``jet-like'', ``monopole-like'', or ``disk-like'', respectively, and then classify magnetospheres by the typical behaviors of their field lines in high latitudes and in low latitudes.  For example a purely monopolar magnetosphere is classified as ``Monopole-Monopole'' while a magnetosphere with field lines that bend upwards in high latitudes and downwards in low latitudes is classified as a ``Jet-Disk'' magnetosphere.  The classification of a magnetosphere is accomplished by subjective inspection; as such the boundary between high and low latitudes and what is more ``monopolar'' than not varies from magnetosphere to magnetosphere.  

\begin{figure*}[p]
\centering
     \includegraphics[height=0.7\textheight,keepaspectratio]{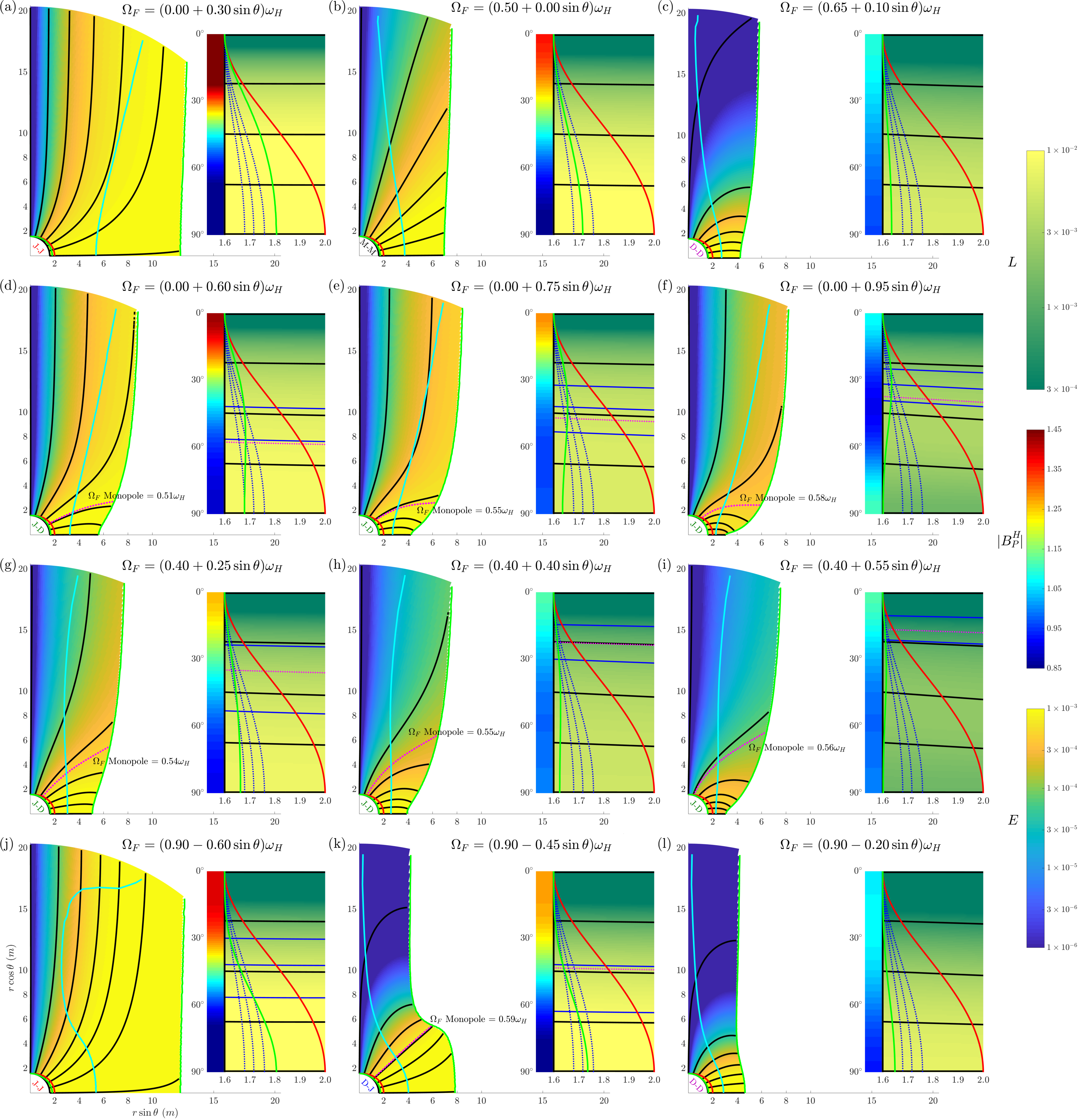}
     \caption{Twelve different magnetospheres displayed using the same conventions as in Figures \ref{Fig:Figure1} and \ref{Fig:Figure2}. The inner and outer light surfaces are shown in green, the boundary of the ergosphere is marked in red, the separation surface is shown in cyan, and the most monopolar field line (if relevant) is marked in dotted magenta.  The magnetosphere classification type (Jet, Monopole, Disk) of high and low latitude regions is denoted by text inside the horizon.  The inset plots show the locations of the inner light surfaces of magnetospheres with uniform field line angular velocities $0.6 \omega_\text{H}$, $0.5 \omega_\text{H}$, and $0.4 \omega_\text{H}$ (furthest from the horizon) as dotted blue lines.  The inset plots on the middle two rows and middle of the bottom row (d, e, f, g, h, i, and k) mark the field lines rotating at $0.6 \omega_\text{H}$, $0.5 \omega_\text{H}$, and $0.4 \omega_\text{H}$ in blue.  All 400 calculated magnetospheres are listed by type in Figure \ref{Fig:Figure4}; any magnetosphere of the same type as one of the 12 above is qualitatively similar in structure.  Monopole-Disk and Disk-Monopole magnetospheres are the only type not shown above as they may be easily imagined as a combination of (a) and (b) or (b) and (c).}
			\label{Fig:Figure3}
\end{figure*}

In Figure \ref{Fig:Figure1} we plot a Jet-Disk magnetosphere with horizon field line angular velocity distribution $\Omega_\text{F} = (0.20 + 0.65 \sin \theta) \omega_\text{H}$.  The magnetosphere is limited to the region interior to the outer light surface, shown as a green line.  The separation surface is shown as a cyan line and might be considered as a more realistic outer boundary, as it delineates the region where the forces on the plasma shift from being dominated by outward centripetal forces to being dominated by inward gravitational forces.  As such a large accumulation of plasma might be expected near the separation surface, breaking the assumption of a force-free plasma and the rigid conservation of field-aligned quantities.  In Figure \ref{Fig:Figure2} we plot the same magnetosphere using an $(r, \theta)$ grid focused on the ergoregion.  The shading in Figure \ref{Fig:Figure2} corresponds to the outward momentum flux (or toroidal magnetic field); the strength of the poloidal field on the horizon is shown as a colorbar immediately inside the horizon, allowing for a comparison of the relative strengths of the two magnetic fields.  The inner light surface is shown as a green line while the inner light surfaces of uniformly rotating magnetospheres with field line angular velocities of $0.4 \omega_\text{H}$, $0.5 \omega_\text{H}$, and $0.6 \omega_\text{H}$ are shown as dotted dark blue lines, allowing for a rough determination of correlation of inner light surface location with magnetosphere behavior.  Such correlations and other effects are discussed in more detail below, and an additional twelve magnetospheres of various different types are plotted in Figure \ref{Fig:Figure3} using methods identical to Figures \ref{Fig:Figure1} and \ref{Fig:Figure2}.         

\begin{figure*}[ht]
\centering
     \includegraphics[width=0.8\textwidth]{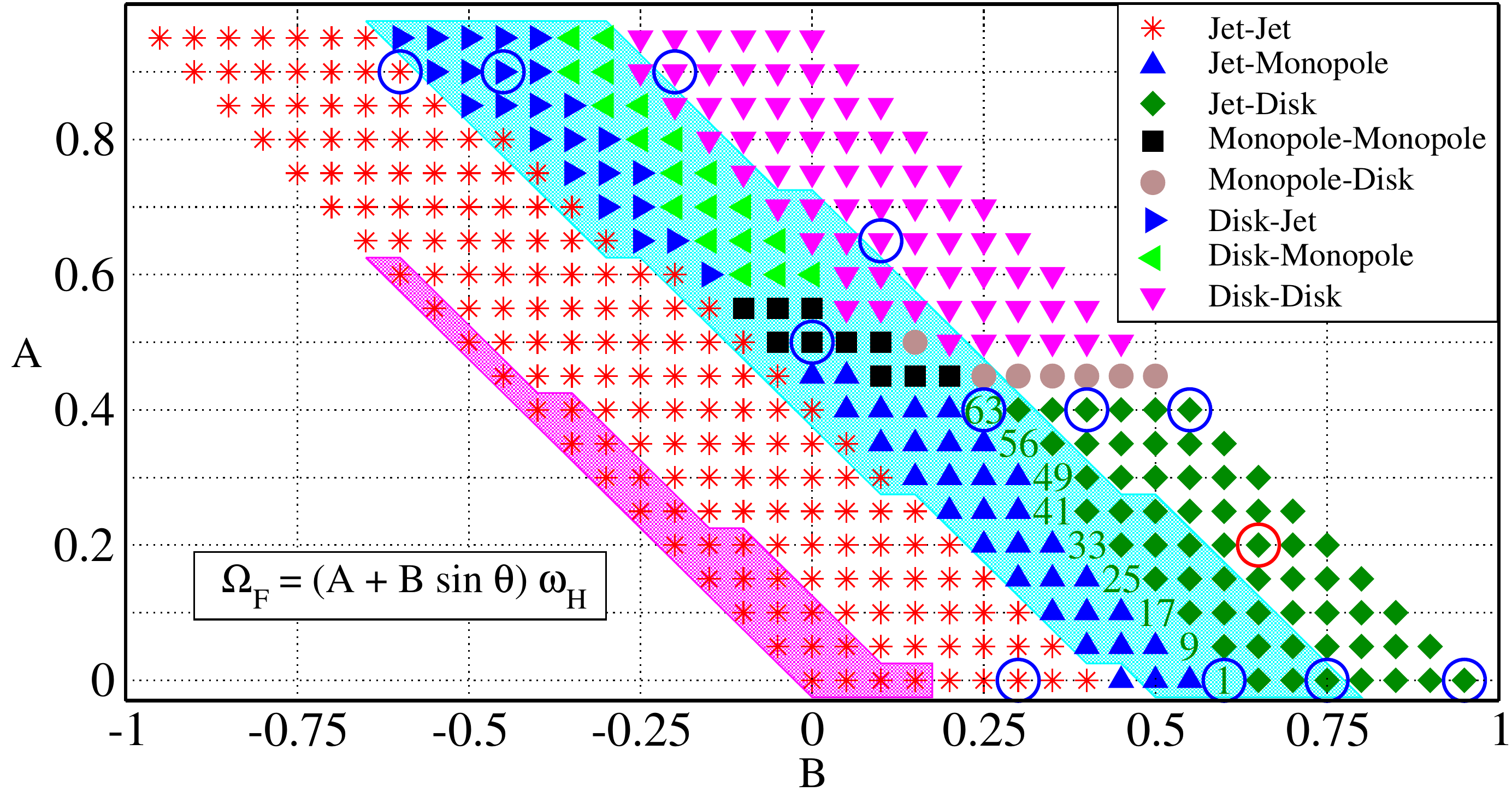}
     \caption{Classification of magnetospheres according to their high and low latitude behaviors as a function of field line angular velocity on the horizon.  The boundaries between regions are moderately susceptible to subjective interpretation, but the classification within a region is robust.  The cyan shading denotes magnetospheres with at least $90\%$ of the maximum luminosity (cf. Figure \ref{Fig:Figure5}); the magenta shading denotes magnetospheres with at least $90\%$ of the maximum rate of angular momentum extraction (cf. Figure \ref{Fig:Figure6}).  The green numbers denote the Jet-Disk magnetosphere numbering scheme used in Section \ref{Sec:JDBehavior}.  The blue circles mark the magnetospheres shown in Figure \ref{Fig:Figure3}; the red circle marks the magnetosphere shown in Figures \ref{Fig:Figure1} and \ref{Fig:Figure2}.}  
	   \label{Fig:Figure4}
\end{figure*}

Monopole-Jet magnetospheres were the only type not found.  The other eight types were present, although as shown in Figure \ref{Fig:Figure4} some types were much more common than others.  By far the most common type was Jet-Jet, followed by Jet-Disk and Disk-Disk.  The boundaries and transitions between different types of magnetospheres are slightly fuzzy due to the subjective nature of their classification, but within a given classification region the behaviors are robust. 

As a general rule the structure of a magnetosphere is predictable by considering the average field line angular velocity of small collections of field lines.  If the average is less than half of the horizon's angular velocity, $\langle\Omega_\text{F} \rangle < 0.5 \omega_\text{H}$, then the field lines will bend upwards toward the azimuthal axis.  If the average is greater than half of the horizon's angular velocity, $\langle \Omega_\text{F} \rangle > 0.5 \omega_\text{H}$, then the field lines will bend towards the equatorial plane.  The strength of either bending increases the further away from $0.5 \omega_\text{H}$ the average field line angular velocity becomes.  

The exception to the above rule is when contradictory preferences are present, such as when high latitude groups want to bend downwards and low latitude groups want to bend upwards.  In that case one group will generally dominate over the other and cause the entire magnetosphere to be either completely Jet-Jet or Disk-Disk.  However there are a few transitional magnetospheres where neither behavior dominates, resulting in Disk-Jet magnetospheres where field lines converge along an approximately $45^\degree$ line through the poloidal plane (as in the middle of the bottom row of Figure \ref{Fig:Figure3}). 

The structure of the magnetospheres near the horizon and inside the ergoregion are shown using an ($r$, $\theta$) grid in Figure \ref{Fig:Figure2} and in the inset plots of Figure \ref{Fig:Figure3}.  As might be expected from monopolar boundary conditions and a minimum energy solution, no significant bending of field lines occurs within the ergoregion; this means that the distribution of horizon field line angular velocity is highly predictive of the shape and location of the inner light surface.  This in turn means that the bending of field lines and structure of a magnetosphere can just as easily be attributed to inner light surface location as to average field line angular velocity.  In other words it can be said that field lines want to bend upwards when they cross the inner light surface closer to the outer limits of the ergoregion and want to bend downwards when they cross the inner light surface near the horizon.      

There is a third potential indicator of magnetosphere structure in addition to average field line angular velocity and the location of the inner light surface.  The colorbars to the left of the horizon in Figure \ref{Fig:Figure2} and the inset plots of Figure \ref{Fig:Figure3} are measures of poloidal magnetic field strength, while the shading outside the horizon is a measure of the toroidal magnetic field (as the conserved angular momentum Poynting flux).  Comparison of the two indicates that large-scale field line bending could also be predicted via the relative strengths of poloidal and toroidal magnetic fields on the horizon.  A strong toroidal field relative to the poloidal field generally causes bending towards the azimuthal axis, while a weak toroidal field and stronger poloidal field results in bending towards the equatorial plane.
 
In Jet-Disk and Disk-Jet magnetospheres a single monopolar field line can be defined as the separatrix between the two regions of opposite bending.  We determined that separatrix by doing a Cartesian $(x,z)$ transformation from the $(r, \theta)$ computational grid near the outer boundary of the magnetosphere followed by finding the absolute minimum of the second derivative in $x$ along different field lines.  The resulting monopolar field line was then visually verified in comparison with the entire magnetosphere to ensure reasonableness.  In general the field line angular velocity of that field line falls between $0.5 \omega_\text{H}$ and $0.6 \omega_\text{H}$, compatible with the notion that $0.5 \omega_\text{H}$ field lines ``want'' to be straight.  Not much more than compatibility should be concluded, however, as a careful inspection of the monopolar field lines drawn in magenta in Figure \ref{Fig:Figure3} makes it clear that the determination of monopolarity can be somewhat arbitrary and dependent upon the region of the magnetosphere chosen for analysis.

Jet-Disk magnetospheres are perhaps the most interesting here, due to both the presence of a clear separatrix between magnetosphere regions and a greater degree of astrophysical viability than Disk-Jet magnetospheres.  Jet-Disk magnetospheres have two characteristic attributes.  First, the field line angular velocity on the azimuthal axis must be less than or equal to $0.4 \omega_\text{H}$.  Second, the field line angular velocity must increase from the azimuthal axis to the equatorial plane with an ultimate value greater than or equal to $0.6 \omega_\text{H}$.  This is again compatible with the general rule of low/high $\Omega_\text{F}$ jet/disk bending, and the more extreme the difference between azimuthal axis and equatorial plane field line angular velocities the more obvious the ``jet-disk'' behavior becomes.  There is significant variation in the amount of energy and angular momentum flowing into either the ``jet'' or to the ``disk'', as shown by the shading in the middle two rows of Figure \ref{Fig:Figure3}; we explore that variation in more detail below in Section \ref{Sec:JDBehavior}.    


\subsection{Energy and Angular Momentum Extraction} \label{Sec:EandL}

\begin{figure}[ht]
\centering
     \includegraphics[width=0.8\columnwidth]{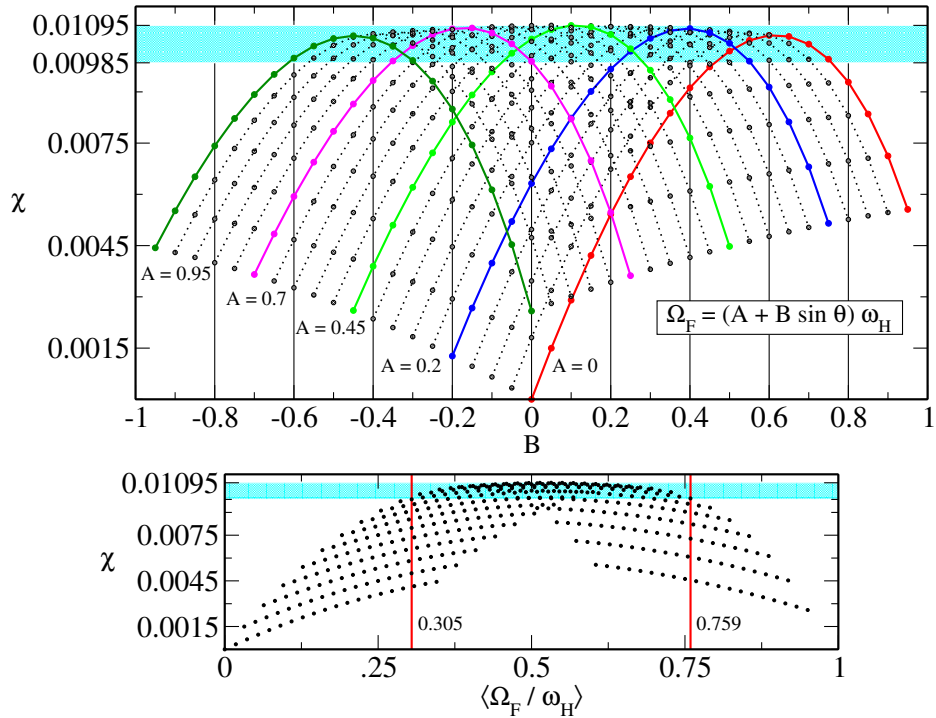}
     \caption{The net rate of black hole energy extraction for all magnetospheres in terms of the dimensionless parameter $\chi$.  The top panel plots lines of constant $A$; the bottom panel plots $\chi$ as a function of average field line angular velocity on the horizon.  The cyan shading denotes the region within $90\%$ of the maximum luminosity.  That region includes every type of observed magnetosphere, as shown by the compatible cyan shading in Figure \ref{Fig:Figure4}.}  
	   \label{Fig:Figure5}
\end{figure}

We measure the net rate of black hole energy extraction via the dimensionless parameter $\chi$, calculated as an integral over the horizon (cf. \citep{LGATN2014}):
\begin{equation} \label{Eqn:ChiEqn}
\chi =  \frac{1}{2} \frac{a_*^2}{\left(r_{+*}^2 + a_*^2 \right)} \int_0^\pi Q \left(1 - Q \right) \frac{A_{\phi, \theta}^2 \sin \theta}{r_{+*}^2 + a_*^2 \cos^2 \theta} d \theta. 
\end{equation} 
Here $Q$ is a unitless scaling of the field line angular velocity on the horizon; $Q = A + B \sin \theta$.  In terms of $\chi$, the net luminosity is given by:
\begin{align}
P &= \int_{r+} T^r{}_t \sqrt{-g} d \theta d \phi \nonumber \\
 &= 6.5 \times 10^{20} \cdot \chi \cdot r_{x*}^4 \frac{B_x^2}{\text{G}^2} \frac{m^2}{M_\odot^2}  \frac{\text{erg}}{\text{s}}.
\end{align} 
Here $a_*$ and $r_{+*}$ are dimensionless measures of black hole spin and horizon radius; $a = a_* m$ and $r_\text{H} = r_{\text{H}*} m$.  The quantity $B_x$ corresponds to monopolar magnetic field strength at dimensionless radius $r_{x*}$, in the sense that in the Newtonian limit of a monopole we would have a magnetic field that in spherical orthonormal coordinates is given by: 
\begin{equation}
\mathbf{B} = \frac{B_x r_{x}^2}{r^2} \hat{r}.
\end{equation} 
We emphasize that the definitions of $B_x$ and $r_{x*}$ are made purely for convenient compatibility with our monopolar boundary conditions and should be taken to be nothing more than a rough average of magnetic field strength as our magnetospheres are neither Newtonian nor generally truly monopolar.  The rate at which a given magnetosphere extracts energy in terms of $\chi$ is shown in Figure \ref{Fig:Figure5}.

\begin{figure}[ht]
\centering
     \includegraphics[width=0.8\columnwidth,clip=true]{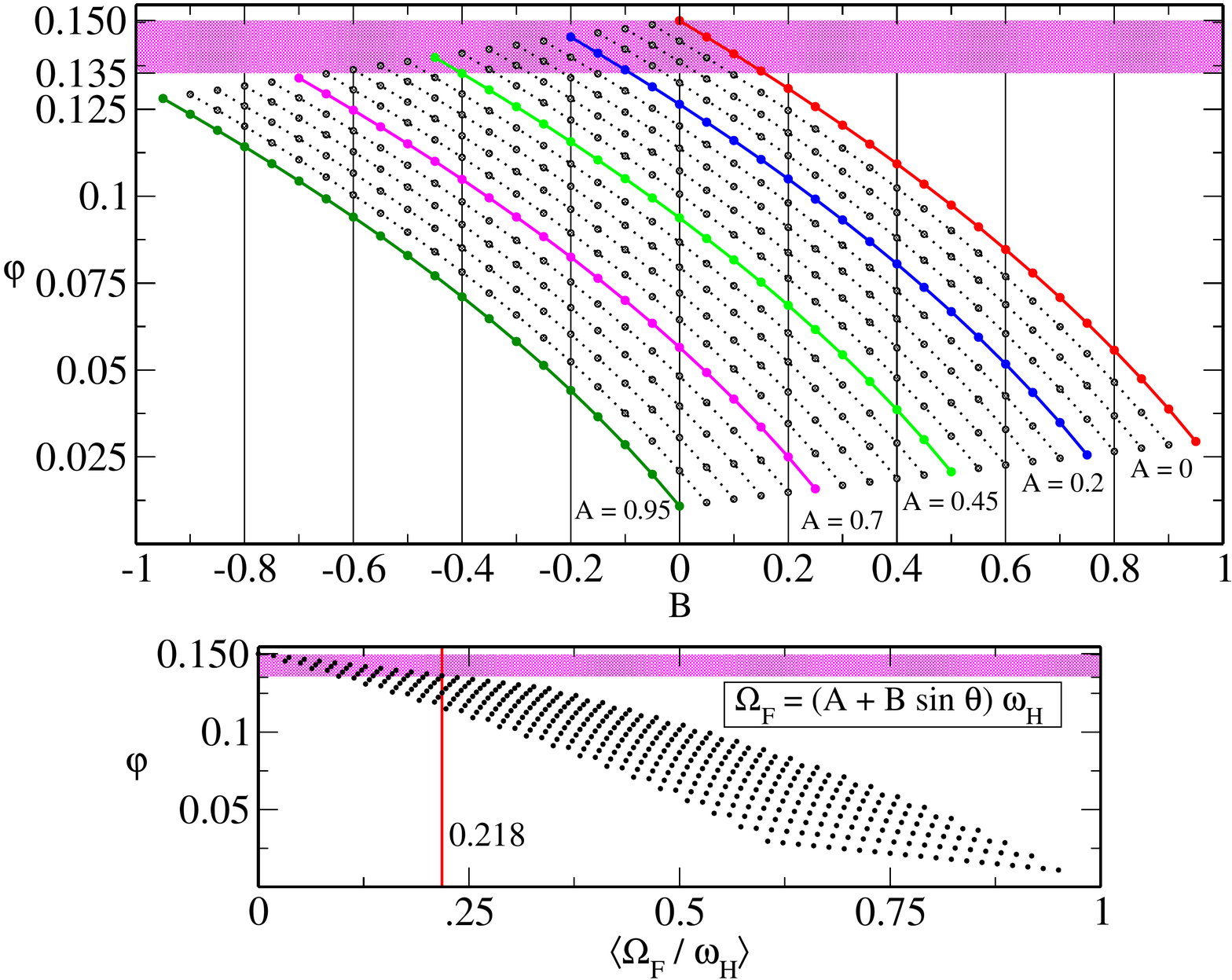}
     \caption{The net rate of black hole angular momentum extraction for all magnetospheres in terms of the dimensionless parameter $\varphi$.  The top panel plots lines of constant $A$; the bottom panel plots $\varphi$ as a function of average field line angular velocity on the horizon.  The magenta shading denotes the region within $90\%$ of the maximum rate of angular momentum extraction.  That region is entirely composed of Jet-Jet magnetospheres, as shown by the compatible magenta shading in Figure \ref{Fig:Figure4}.}  
	   \label{Fig:Figure6}
\end{figure}

We measure the net rate of angular momentum extraction via the dimensionless parameter $\varphi$ in almost identical fashion to the measurement of $\chi$:  
\begin{equation}
\varphi =  \frac{1}{2} a_* \int_0^\pi  \left(1 - Q \right) \frac{A_{\phi, \theta}^2 \sin \theta}{r_{+*}^2 + a_*^2 \cos^2 \theta} d \theta.
\end{equation}
In terms of $\varphi$, the net rate of black hole angular momentum extraction is given by:  
\begin{align}
K &= -\int_{r+} T^r{}_\phi \sqrt{-g} d \theta d \phi \nonumber \\
&= 3.2 \times 10^{15} \cdot \varphi \cdot r_{x*}^4 \frac{B_x^2}{\text{G}^2} \frac{m^3}{M_\odot^3} \text{erg}. 
\end{align} 
The rate at which a given magnetosphere extracts momentum in terms of $\varphi$ is shown in Figure \ref{Fig:Figure6}.

Perhaps the most striking feature of the net rate of energy extraction in Figure \ref{Fig:Figure5} is the very broad peak.  The maximum luminosity corresponds to a horizon field line angular velocity of $\Omega_\text{F} = (0.5 + 0.05 \sin \theta) \omega_\text{H}$, but there are a large number of magnetospheres that have effectively equivalent luminosities within $90\%$ of that maximum.  Those magnetospheres encompass every single type of observed magnetosphere, as shown by the cyan shading in Figure \ref{Fig:Figure4}.  The bottom panel of Figure \ref{Fig:Figure5} indicates that average field line angular velocity is not very predictive of a maximally luminous magnetosphere; anything from $\langle \Omega_\text{F} \rangle = 0.3 \omega_\text{H}$ to $\langle \Omega_\text{F} \rangle = 0.8 \omega_\text{H}$ can yield a very close to maximum luminosity magnetosphere.  Averages near $0.5 \omega_\text{H}$ could be assumed to be relatively luminous and averages closer to the outer limits of energy extracting magnetospheres ($\langle \Omega_\text{F} \rangle = 0$ or $\langle \Omega_\text{F} \rangle = \omega_\text{H}$) could be assumed to be relatively dim, but anything else would require closer analysis.

The monopolar magnetosphere structure explored in \citep{BZ77} (in the present language $A = 0.5$, $B = 0$, and $A_\phi = \cos \theta$) corresponds to a $\chi$ parameter of:
\begin{equation} \label{Eqn:BZEnergyEqn}
\chi_{\text{BZ77}} = \frac{1}{8} \frac{a_*^2}{r_{+*}^2 + a_*^2} \int_0^\pi \frac{\sin^3 \theta}{r_{+*}^2 + a_*^2 \cos^2 \theta} d \theta.
\end{equation}    
In this work we used a dimensionless black hole spin parameter $a_* = 0.8$ and corresponding horizon radius $r_{+*} = 1.6$, yielding $\chi_{\text{BZ77}} = 0.0124$.  Despite being well outside the ``low spin'' assumption used in \citep{BZ77}, it is still only about $10\%$ over our maximum value of $\chi$.  This is primarily a result of our solutions concentrating more horizon poloidal magnetic flux near the azimuthal axis than a monopole would, as shown by the middle of the top row of Figure \ref{Fig:Figure3}.  

The closeness of such a crude approximation indicates that almost any estimate of the net luminosity of an energy-extracting black hole magnetosphere based in some way on the assumptions of a Blandford and Znajek monopole as in \citep{MG2004, TchekhovskoyNarayanMcKinney2010,PanYu2015} will in general be successful.  There are a wide range of magnetospheres within $10\%$ or so of the near energy maximum such an estimate would yield, including essentially every magnetosphere with an average horizon field line angular velocity $\langle \Omega_\text{F} \rangle \approx 0.5 \omega_\text{H}$ typically used in such estimates.  This implies that outside of more extreme distributions of field line angular velocity noticeable discrepancies might only be expected to emerge if finer detail (i.e. magnetosphere structure) is considered.          

The net rate of angular momentum extraction is more discriminating in its behavior than the net rate of energy extraction, as shown in Figure \ref{Fig:Figure6}.  Very low field line angular velocity magnetospheres always extract more angular momentum than others.  Those magnetospheres also always correspond to Jet-Jet magnetospheres, as shown by the magenta shading in Figure \ref{Fig:Figure4}.  The spread in the rate of angular momentum extraction for a given average horizon field line angular velocity is typically not very large, so it is also generally safe to assume that if average field line angular velocity is increased then the net rate of angular momentum extraction will be decreased. 

So far we have only considered net rates of energy and angular momentum extraction.  However the direction of those flows could in many instances be of far greater importance than the net values, as indicated by comparing the luminosities of Figure \ref{Fig:Figure5} with magnetosphere type in Figure \ref{Fig:Figure4} and magnetosphere structure in Figure \ref{Fig:Figure3}.  Magnetospheres near the luminosity maximum are those that are closest to exhibiting monopolar behaviors; magnetospheres that bend most tightly towards either the azimuthal axis or the equatorial plane are the dimmest.  When compared with the rates of angular momentum extraction, we find that magnetospheres that would most rapidly spin down a black hole will be fairly underluminous in a global sense, but that most of that energy will be very tightly directed along the azimuthal axis.  Magnetospheres that would take the longest amount of time to spin down a black hole will be similarly underluminous in a global sense, but most of that energy will be transmitted into a small nearby region in the equatorial plane.  So while some magnetospheres might be dimmer overall than others, they might direct that energy in a far more efficient fashion to specific regions of interest while at the same time correlating with other behaviors due to their different rates of angular momentum extraction.

We examine the importance of the direction of energy and angular momentum flows in more depth in the next section within the context of Jet-Disk magnetospheres.  


\subsection{Structure of Jet-Disk Magnetospheres} \label{Sec:JDBehavior}


In this section we divide the Poynting fluxes of energy and angular momentum of Jet-Disk magnetospheres into their ``jet'' and ``disk'' components and compare their magnitudes.  Jet-Disk disk magnetospheres could be one of the more interesting types of magnetosphere we found, as they most closely resemble what might be expected in an astrophysical environment: open field lines aligned with the azimuthal axis and lower latitude structures compatible with a connection to nearby accreting matter.  That is not to say that they are of exclusive interest, but rather that within a relatively simple set of assumptions they might be able to describe (or be compatible with) both jet-like structures and accretion structures.  The Jet-Disk magnetospheres we found are also unique in that they allow for a clear delineation between the ``jet'' region and the ``disk'' region; that delineation allows us to move beyond the net rates of energy and angular momentum extraction explored in the previous section without having to more arbitrarily subdivide magnetospheres.  

Of the 400 calculated magnetospheres, 69 are classified as Jet-Disk, with $A$ parameters in the distribution of horizon field line angular velocity $\Omega_\text{F} = (A + B \sin \theta) \omega_\text{H}$ ranging from $0.0$ to $0.4$.  In order to discuss the general tendencies of Jet-Disk magnetospheres we first number each of the 69 magnetospheres by their different $A$ parameters as shown in Figure \ref{Fig:Figure4}; magnetospheres with $A = 0.0$ and $B = [0.60 \ldots 0.95]$ are numbered by increasing $B$ from 1 to 8, those with $A = 0.05$ and $B = [0.55 \ldots 0.90]$ are numbered by increasing $B$ from 9 to 16, and so on.   

We separate magnetospheres into jet and disk regions by using the most monopolar field line (i.e. the line with minimal bending in the poloidal plane) as the separatrix between the two regions.  The determination of the most monopolar field line is slightly arbitrary in that it is dependent upon the parameters and regions used to measure bending, which in Section \ref{Sec:GeneralStructure} led to some variability in the field line angular velocity of the most monopolar field line.  In this case that variability is not significant, as both the ratios of energy and angular momentum flow to jet and disk regions as well as the trends in their magnitudes from magnetosphere to magnetosphere do not significantly change over the spread of what could reasonably be called a valid separatrix between jet and disk regions. 

\begin{figure}[ht]
\centering
     \includegraphics[width=0.8\columnwidth]{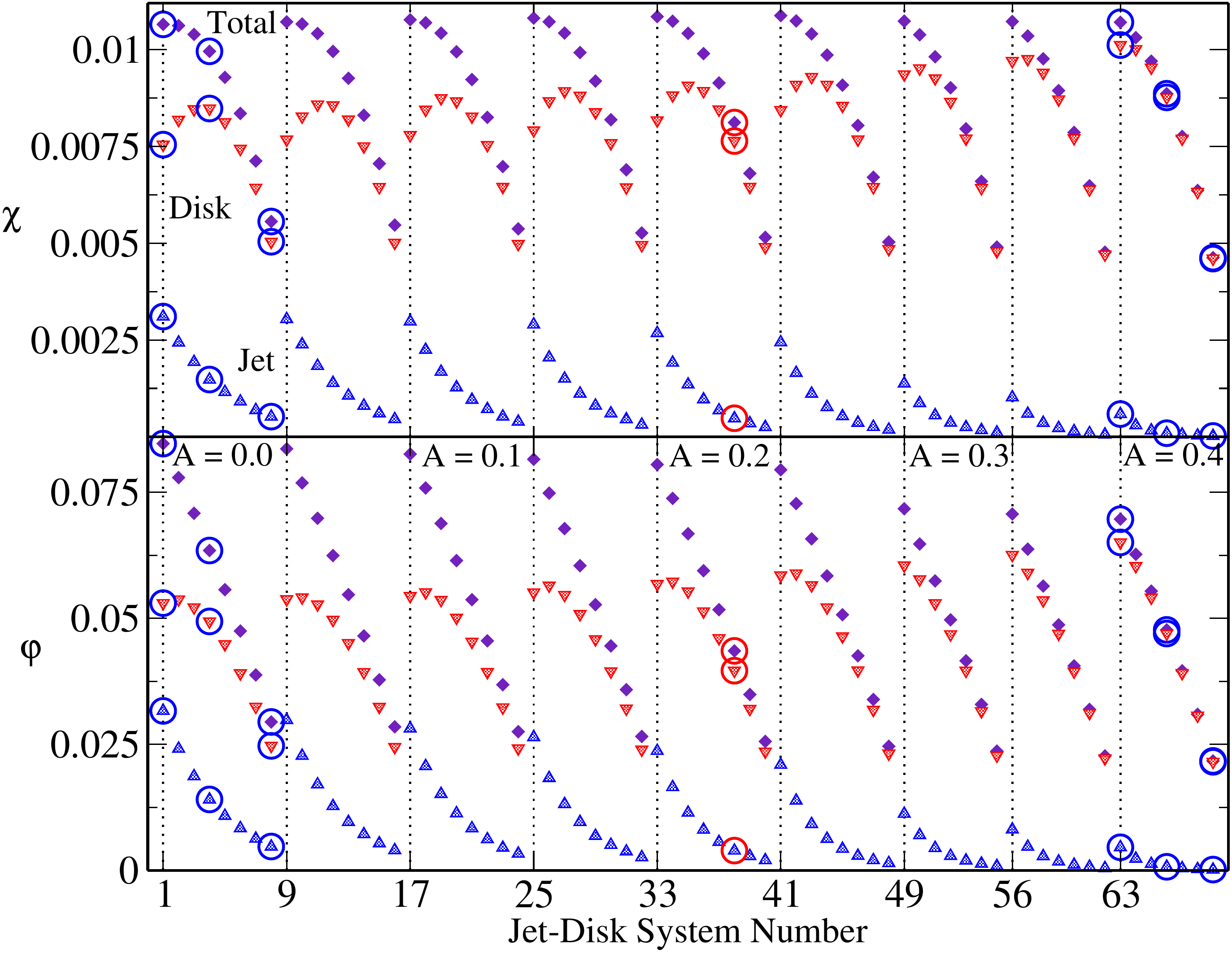}
     \caption{The rates of energy and angular momentum flow into the jet and disk regions for all 69 Jet-Disk magnetospheres.  The magnetosphere numbering scheme is shown in Figure \ref{Fig:Figure4}; the vertical lines separate magnetospheres into $A$ parameter sections, while $B$ parameter increases across a section ($\Omega_\text{F} = (A + B \sin \theta) \omega_\text{H}$).  The total rates of energy and angular momentum extracted are fairly constant but the ratio between jet and disk regions can be large.  The maximum rate of energy flow into the jet is $\chi_\text{Jet Max} \approx 3 \times 10^{-3}$ while the minimum is $\chi_\text{Jet Min} \approx 2 \times 10^{-5}$, around 130 times smaller.  The blue circles mark the magnetospheres shown in Figure \ref{Fig:Figure3}; the red circle marks the magnetosphere shown in Figures \ref{Fig:Figure1} and \ref{Fig:Figure2}.}  
	   \label{Fig:Figure7}
\end{figure}

Figure \ref{Fig:Figure5} shows the rates of energy and angular momentum flux into both jet and disk regions for all 69 Jet-Disk magnetospheres, grouped by the $A$ parameter of their horizon field line angular velocities.  In general much more energy flows into the disk region than into the jet region.  Similarly much more angular momentum flows into the disk region than into the jet region, although the relative difference is generally smaller.  The total rates of energy and angular momentum extraction don't vary that much across all of the Jet-Disk magnetospheres, but as the average field line angular velocity goes up very little of either gets sent into the jet region.   

\begin{figure}[ht]
\centering
     \includegraphics[width=0.8\columnwidth]{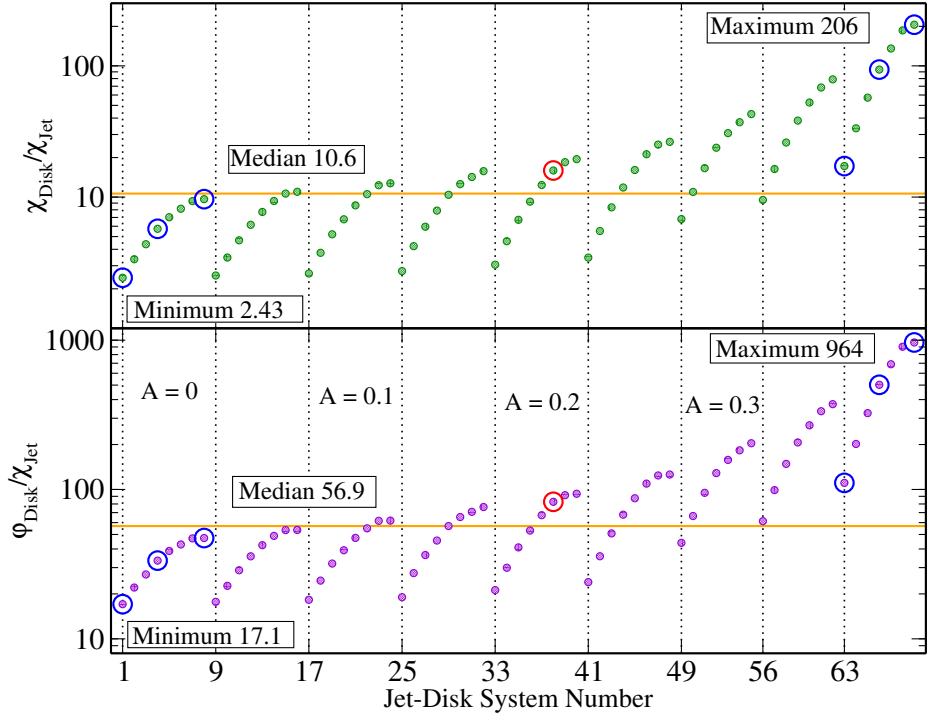}
     \caption{The amount of energy and angular momentum flowing into the disk region per unit of energy flowing into the jet region in Jet-Disk magnetospheres.  The magnetosphere numbering scheme is shown in Figure \ref{Fig:Figure4} (the same as used in Figure \ref{Fig:Figure5}).  For any assumed energy flow into the jet-like structure aligned with the azimuthal axis the concurrent amounts of energy and angular momentum flowing towards the equatorial plane can vary widely.  The blue circles mark the magnetospheres shown in Figure \ref{Fig:Figure3}; the red circle marks the magnetosphere shown in Figures \ref{Fig:Figure1} and \ref{Fig:Figure2}.}  
	   \label{Fig:Figure8}
\end{figure}

The amounts of energy and angular momentum flowing into the disk region per unit of energy flowing into the jet region are shown in Figure \ref{Fig:Figure8}.  For an assumed jet energy up to 200 times more energy could be flowing into the disk region, with a median of around 10 times more energy.  For that same assumed jet energy there is also a large range of possible momentum fluxes into the disk region; the maximum is almost 60 times larger than the minimum amount.  Those ranges are coupled to the strength of the jet in terms of how sharply field lines bend towards the axis, as illustrated by the two middle rows of Figure \ref{Fig:Figure3}.  The tightest jet-like structures have the smallest ratios between jet energy and momentum and disk energy and momentum, while the loosest jet-like structures have the largest ratios between jet and disk energies and momenta.  

Putting the ratio of jet energy and disk energy and momentum flows into more concrete terms, for every erg of energy flowing into the jet region there will be between $2$ and $200$ ergs of energy flowing into the disk region and between $1\times 10^{-4} m/M_\odot$ and $5 \times 10^{-3} m/M_\odot$ erg-seconds of angular momentum flowing into the disk region, with median values given by:
\begin{align}
\text{1 Jet erg} &\sim \text{10 Disk erg} \nonumber \\
	&\sim \text{$3 \times 10^{-4} \frac{m}{M_\odot}$ Disk erg-s}.
\end{align}
Potential implications of those ranges are discussed below in Section \ref{Sec:JetDiskDiscussion}.     


\section{Discussion} \label{Sec:Discussion}

In this section we discuss two topics in more depth.  First we compare our results with our previous work in calculating uniform field line angular velocity magnetospheres.  Second we explore how reasonable the Jet-Disk magnetospheres we found might be, and what potential implications they might have for astrophysical objects.


\subsection{Comparison with Uniform Field Line Angular Velocity}


In our previous numerical work \citep{TTT2017} we calculated the structure of energy extracting force-free black hole magnetospheres as a function of uniform field line angular velocity using the same monopolar boundary conditions along the azimuthal axis and equatorial plane that were used here.  We found that rapidly rotating magnetospheres (referenced to the horizon's angular velocity) with inner light surfaces near the horizon had poloidal magnetic field lines that bent towards the equatorial plane, while slowly rotating magnetospheres with inner light surfaces near the outer boundary of the ergoregion had poloidal magnetic field lines that bent upwards towards the azimuthal axis.  Such behavior may also be seen analytically Appendix \ref{App:PerturbedSolution}, though in a more limited form.

In this work we largely found the same behaviors.  If a collection of field lines had a relatively small average field line angular velocity they bent upwards towards the azimuthal axis; if they had a relatively large average field line angular velocity they bent downwards towards the equatorial plane.  The only exceptions to that behavior were in cases where adjacent groups had incompatible bending preferences; in those cases the group with most extreme field line angular velocity would generally ``win'' and bend the other group.    

Our previous work speculated that consideration of the direction of energy and angular momentum flows might be critical to any consideration of black hole energy extraction as a plausible central engine driving astrophysical phenomena.  In that speculation we were hampered by the crudity of our assumptions, perhaps most notably that of uniform field line angular velocity, but nonetheless were able to use the timescale of a transient object as an example of how the direction of energy and angular momentum flows could potentially modify observed behaviors.  

Having now solved for more realistic distributions of magnetic field line angular velocity we are more strongly convinced that the direction of energy flow should be a primary consideration in determining the applicability of black hole energy extraction to any given astrophysical object.  Within the context of Jet-Disk magnetospheres, for example, any assumed amount of ``jet'' energy extracted from the rotating black hole might need to be coupled to a consideration of the effects of the concurrent flows of energy and angular momentum into nearby accreting matter, something we discuss in more detail below in Section \ref{Sec:JetDiskDiscussion}.

Lastly, our previous work speculated that changes in magnetosphere structure could be more significant in varying luminosity than changing either black hole spin or magnetic field strength.  That is what we found here; within Jet-Disk magnetospheres we found that jet energies varied by a factor of $130$ (Figure \ref{Fig:Figure7}).  As a general rule the rate of energy extraction varies with black hole spin as $a^2$ (e.g. Equation \ref{Eqn:BZEnergyEqn} or \ref{Eqn:ChiEqn}), although for very high spin better approximations can be made \citep{TchekhovskoyNarayanMcKinney2010}.  If we select what might be a reasonable range of black hole spins for active galactic nuclei, $0.3m$ to $0.95m$ \citep{VolonteriSikoraLasota2007}, then an $a^2$ estimate only yields a factor of $10$ variation in luminosity due to changing black hole spin.  Our previous numerical work \citep{TTT2017} suggests that a factor of $30$ variation might be a better estimate, but that's still sub dominant to the factor of $130$ variation within Jet-Disk magnetospheres found here.  When both effects are combined, jet luminosity variations in excess of a factor of $1000$ could be expected for the jets of Jet-Disk magnetospheres over a reasonable range of black hole spins, a variation that would be correlated with both the degree of jet collimation and the amount of energy and angular momentum concurrently flowing into the disk region. 


\subsection{Jet-Disk Magnetospheres} \label{Sec:JetDiskDiscussion}


In this section we explore some potential implications of the Jet-Disk magnetospheres that were found.  We first examine whether or not the distributions of horizon field line angular velocities leading to those magnetospheres are reasonable, then suggest some restrictions that Jet-Disk magnetospheres might place on black hole energy extraction in astrophysical contexts.

To determine how reasonable the distributions of Jet-Disk horizon field line angular velocities might be, we consider what might be expected of the black hole's nearby environment.  An isolated black hole cannot support a magnetic field, so the magnetic flux that we assume exists on the horizon must be maintained by nearby matter.  A likely configuration of such matter compatible with our assumption of stationarity is matter rotating near the equatorial plane with an angular velocity distribution corresponding to centripetal forces roughly balancing gravitational forces.  For convenience we will call that configuration of matter a ``disk'' (a choice made to aid discussion, not to imply preference for a thin/thick disk, torus, or other structure).  The disk is likely to be highly conductive, meaning that the magnetic field will rotate with the disk and possess a field line angular velocity compatible with the disk's angular velocity.  This means that near the equatorial plane the angular velocity of magnetic field lines should be largest near the black hole and decrease as the distance from the black hole increases, formally vanishing infinitely far away.  

The field lines on the horizon near the equatorial plane should connect to the disk in nearby regions, and by virtue of rotating with the disk should have $\Omega_\text{F} \approx \omega_\text{H}$.  Proceeding up the horizon field lines will connect with the disk further and further away, resulting in a gradual diminishing of field line angular velocity towards the azimuthal axis.  That is exactly the type of horizon field line angular velocity distribution that results in Jet-Disk magnetospheres, and has been used by \citep{YuanBlandfordWilkins2018Arxiv} to calculate magnetospheres that directly connect the horizon to a nearby disk (the problem setup used there prohibits the emergence of open field lines connected to the horizon, however, making the emergence of a Jet-Disk magnetosphere impossible).

In other words the simplest of assumptions one could make about a black hole's environment are both compatible with and intrinsically imply the presence of jet-like structures aligned with the azimuthal axis and disk-like connections near the equatorial plane.  That compatibility indicates that in many contexts asking how a Jet-Disk magnetosphere structure \textit{didn't} form might be a far more difficult question to answer than how it did.

Within the assumption of stationarity, it might be difficult to directly describe the magnetically driven turbulence \citep{BalbusHawley1991} that might drive accretion in a ``disk''.  Nonetheless when horizon field line angular velocity (and magnetosphere structure) is considered in a time-averaged sense, numerical simulations \citep{MG2004} have found Jet-Disk magnetosphere structures and rotation profiles with horizon field line angular velocities ranging from $0.4 \omega_\text{H}$ on the azimuthal axis to $0.8 \omega_\text{H}$ near the equatorial plane.  More recent numerical simulations \citep{NakamuraEtAl2018} have found similar magnetosphere structures, though unfortunately it is uncommon to report magnetosphere rotation profiles. 

If it is assumed that Jet-Disk magnetospheres might correspond to astrophysically relevant magnetospheres, the variations in energy and angular momentum flows found in Section \ref{Sec:JDBehavior} immediately place restrictions on those magnetospheres.  Large fluxes of angular momentum into the ``disk'' might blow away that disk and halt both accretion and black hole energy extraction, implying that there might be a maximum effective ``jet'' luminosity before the jet becomes intrinsically variable.  For any assumed ``jet'' energy there will also be a significant flux of energy that would be absorbed into a relatively compact region of the ``disk'' near the horizon; plasma in that region could then provide a useful reservoir of highly energetic particles from which to launch a true jet in a sheath surrounding the inner Poynting flux jet of a Jet-Disk magnetosphere.   

Quantifying such behaviors in more depth is beyond the scope of this work, and are largely not fundamentally new considerations in any event.  For example, magnetic fields torquing the inner portions of a black hole's accretion disk and modifying its behavior were being considered years before Blandford and Znajek's seminal paper \citep{BZ77} was published \citep{Thorne1974}, and similar effects have been examined before within the context of energy extracting black hole magnetospheres \citep{vanPutten1999, vanPuttenLevinson2003, Li2002}.  What we wish to suggest is not the creation of an entirely new model of energy-extracting black hole magnetospheres, but rather the potentially intrinsic compatibility of near-horizon electromagnetic field structure (as expressed by Jet-Disk distributions of field line angular velocity) with both a compact connection to nearby accreting matter and structures reminiscent of jets.    


\section{Conclusions}


We calculated 400 energy-extracting black hole magnetospheres with varying horizon field line angular velocity distributions given by $\Omega_\text{F} = (A + B \sin \theta) \omega_\text{H}$, corresponding to the first two terms of an expansion of an arbitrary horizon field line angular velocity distribution.     

We found that horizon field line angular velocity and the location of the inner light surface are equally predictive of large scale magnetosphere structure.  Groups of field lines with field line angular velocity $\Omega_\text{F} \lesssim 0.5 \omega_\text{H}$ (inner light surfaces closer to the outer limits of the ergoregion) tend to bend towards the azimuthal axis.  Groups of field lines with field line angular velocity $\Omega_\text{F} \gtrsim 0.5 \omega_\text{H}$ (inner light surfaces closer to the horizon) tend to bend towards the equatorial plane.  The strength of either bending increases the further away from $0.5 \omega_\text{H}$ the field line angular velocity becomes (the closer the inner light surface gets to the horizon or outer limit of the ergoregion).  

We also found that the horizon field line angular velocity distribution perhaps compatible with conditions introduced by nearby accreting matter naturally correspond to magnetospheres that both connect the horizon to that matter and form jet-like structures aligned with the azimuthal axis.  That implies that near-horizon jet launching might in some cases be expected as a general feature of energy-extracting black hole magnetospheres.  Varying black hole spin from $a = 0.3m$ to $a = 0.95m$ coupled to the variations in magnetosphere structure we found could lead to a factor of $1000$ or more difference in the luminosity of such jets.  Much of that variation would be due to changes in magnetosphere structure, both in terms of the degree of jet collimation and in terms of the proportions of energy and angular momentum flowing upwards into the putative jet and nearby accreting matter.


\section*{Acknowledgments}

M.T. was supported by JSPS KAKENHI Grant Number 17K05439, and DAIKO FOUNDATION.  S.T. acknowledges the support of the IPMU and WPI.


\bibliographystyle{ptephy}

\appendix


\section{Perturbed Solution} \label{App:PerturbedSolution}


In this section we study the structure of uniformly rotating black hole magnetospheres analytically by applying perturbation techniques in black hole spin from non-rotating to rotating spacetimes.  Around a non-rotating black hole, there exists a class of exact solutions that are given by:
\begin{align} \label{Eq:SchwarzschildMonopoleSolution}
A_\phi &= B_0 \cos \theta, \nonumber \\
\Omega_\text{F} &= \Omega(\theta), \nonumber \\
\sqrt{-g} F^{\theta r} &= B_0 \Omega(\theta) \sin^2 \theta.
\end{align}
These are a generalization of the solutions originally found by \citep{Michel1973}, and which \citep{BZ77} used (with $\Omega(\theta) = 0$) to find a perturbed monopole solution.  For simplicity we will take the field line angular velocity $\Omega_\text{F}$ to be a constant, given by $\Omega_\text{F} = x (a/4m^2) = x \omega_\text{Hp}$ where $x$ is a unitless weighting factor and $\omega_\text{Hp}$ is taken to correspond to the perturbed angular velocity of the horizon in a rotating spacetime.  Note that $\omega_\text{Hp}$ should not be taken to vanish as $a \rightarrow 0$; rather it should be taken as a convenient weighting of $\Omega_\text{F}$ once an $a \neq 0$ spacetime has been specified.  Once we assume that such a spacetime has been specified, then to second order in spin the fields have corrections that are given by:
\begin{align}
A_\phi &= B_0 \cos \theta - \frac{a^2}{16 m^4} \frac{B_0}{9 m} \left(x - \frac{1}{2}\right) \left(r - 2m \right) \left(44 r^2 + 13 m r + 14 m^2 \right) \cos \theta \sin^2 \theta \nonumber \\
&+ a^2 B_0 R_\text{Corr} \cos \theta \sin^2 \theta, \nonumber \\
\Omega_\text{F} &= x \omega_\text{Hp}, \nonumber \\
\sqrt{-g} F^{\theta r}  &= \frac{1}{B_0} \left(x \omega_\text{Hp} - \frac{a}{4m^2}\right) \left(B_0^2 - A_\phi^2 \right),
\end{align}
where:
\begin{align}
R_\text{Corr} &= -\frac{1}{m^4} \left[\frac{m^2 + 3mr - 6r^2}{12} \ln \left(\frac{r}{2m} \right) + \frac{11 m^2}{72} +  \frac{m^3}{3r} + \frac{mr}{2} - \frac{r^2}{2} \right] \nonumber \\
&- \left(\frac{2r^3 - 3mr^2}{8m^5} \right) \left[\text{Li}_2 \left(\frac{2m}{r}\right) - \ln \left(1 - \frac{2m}{r} \right) \ln \left(\frac{r}{2m} \right) \right].
\end{align}  
Here $\text{Li}_2$ is the dilogarithm, defined as:
\begin{equation}
\text{Li}_2 (x) = \int_x^0 \frac{1}{t} \ln \left(1 -t \right) dt.
\end{equation}
\begin{figure}[!h]
\centering
     \includegraphics[width=0.9\textwidth]{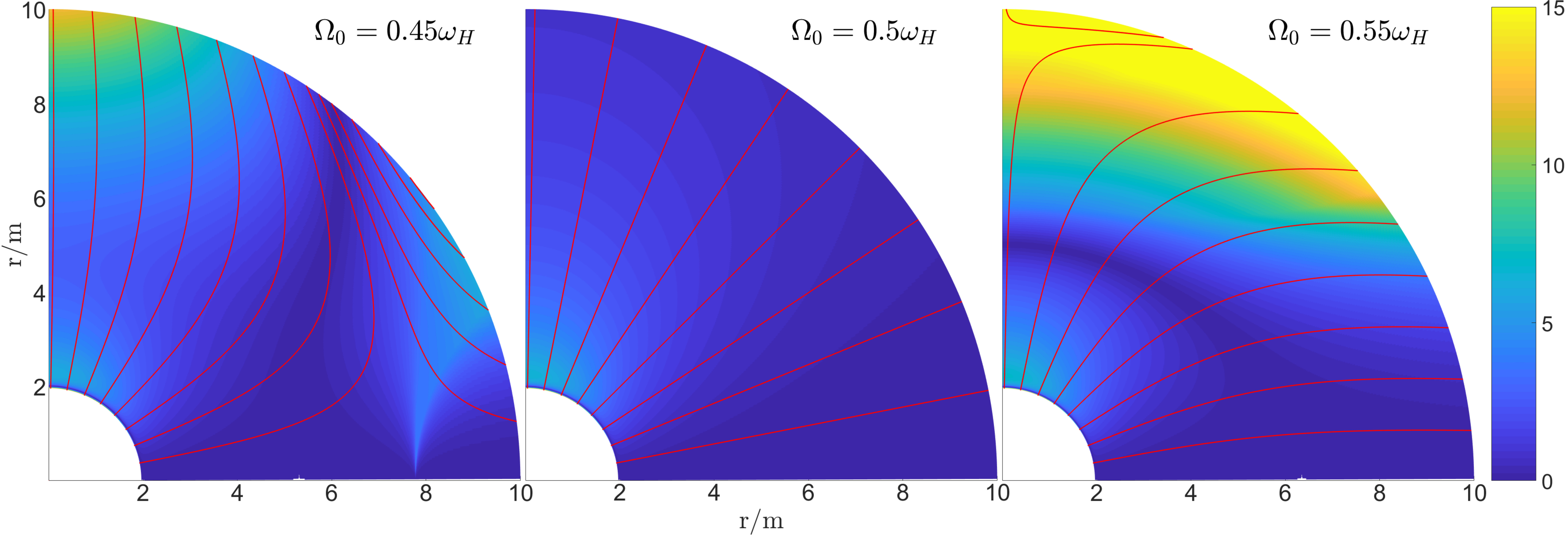}
     \caption[Perturbed Rotating Monopoles]{Three magnetospheres with $\Omega_0 = 0.45 \omega_\text{Hp}$, $\Omega_0 = 0.5 \omega_\text{Hp}$, $\Omega_0 = 0.55 \omega_\text{Hp}$ (where $\omega_\text{Hp} \equiv a / 4m^2$) for black hole spin $a = 0.3m$.  The background shading is the percent error of the solutions \citep{TTT2017}.  The $\Omega_0 = 0.5 \omega_\text{H}$ solution is a separatrix between two classes of solutions that can exhibit significant modifications to the structure of the poloidal field when extended from Schwarzschild to Kerr spacetimes.  We have deliberately chosen to extend the domain to include topological changes to the field (i.e. divergences from monopolarity).  In practice those regions are mostly indicative of a breakdown in the solution, and should be viewed with some suspicion.}  
	   \label{Fig:PerturbedMonopole}
\end{figure}
In the above the single perturbed monopole solution found by \citep{BZ77} is recovered when $x \rightarrow 1/2$, resulting in the well-known result that monopolar fields rotate at roughly half the field line angular velocity of the horizon.  The correction to $A_\phi$ that is a function of $x$ is found by demanding that the fields remain monopolar when $r = 2m$ and acknowledging that significant changes to the fields may occur in more distant regions.  Formally that correction diverges as $r \rightarrow \infty$, but as we have argued in the main text demanding a smooth extension from the horizon to spatial infinity is ill-advised in general, though it might be appropriate in some contexts.  The behavior of the field lines for different values of $x$ are shown in Figure \ref{Fig:PerturbedMonopole} (suppressing the common $R_\text{corr}$ term, as it is largely irrelevant to structure and dies out as $1/r$).  It is immediately apparent that when $x \lesssim 0.5$ the field lines bend upwards and when $x \gtrsim 0.5$ the field lines bend downwards, compatible with our numerical results.  For any domain with finite outer radius there exists a range of solutions $x = 1/2 \pm \epsilon$ that exhibit the same bending behaviors while maintaining error comparable to the well-known perturbed monopole solution found by \citep{BZ77}.

We are not the first to suggest a general rule coupling magnetosphere bending behaviors to field line rotation.  Impedance matching arguments can be made using resistive membranes on the horizon and at spatial infinity (to include surfaces approximating spatial infinity) to suggest that field lines with diverging angular separation should have $\Omega_\text{F} \gtrsim 0.5 \omega_\text{H}$ and that field lines with converging angular separation should have $\Omega_\text{F} \lesssim 0.5 \omega_\text{H}$ \citep{Penna2015}.  In other words finding $\Omega_\text{F} \sim 0.5 \omega_\text{H}$ as a separatrix between magnetosphere bending behaviors as shown here (with a monopolar configuration coinciding with the separatrix) is not a very surprising result; the primary unknown question is what bending behaviors should be expected.


\end{document}